\documentclass{IEEEtran}
\usepackage[table]{xcolor}
\usepackage{cite}
\usepackage{graphicx}
\usepackage{lipsum}
\usepackage{stfloats}
\usepackage{amssymb}
\usepackage{pifont}
\usepackage{bbding}
\usepackage{booktabs}
\usepackage{ctable}
\usepackage{threeparttable}
\usepackage{footmisc}
\usepackage{framed} 
\usepackage{array}
\usepackage{multirow}
\usepackage{longtable}
\usepackage{booktabs}
\usepackage{rotating}
\usepackage{fancyhdr}
\usepackage{lastpage}
\usepackage{pifont}
\usepackage{layout}
\usepackage{wrapfig}
\usepackage{xcolor}
\usepackage{tabularx}
\usepackage{tabu}
\usepackage{enumitem}
\usepackage{url}

\graphicspath{{Figures/}}

\usepackage{array}

\DeclareGraphicsExtensions{.png, .eps}

\begin{document}
	\title{The Syntactic–Semantic Internet: 
Engineering Infrastructures for Autonomous Systems
}
	\author{
 Mallik Tatipamula~\IEEEmembership{Fellow,~IEEE,}
 Xuesong Liu,~\IEEEmembership{Student Member,~IEEE,} Yao Sun,~\IEEEmembership{Senior Member,~IEEE,} and Muhammad Ali Imran~\IEEEmembership{Fellow,~IEEE}
	\thanks{
Mallik Tatipamula is with Ericsson Silicon Valley, United States;
Xuesong Liu, Yao Sun and Muhammad Ali Imran are with University of Glasgow, United Kingdom.
	
	}}	
	\maketitle
 

\begin{abstract}
The Internet has evolved through successive architectural abstractions that enabled unprecedented scale, interoperability, and innovation. Packet-based networking enabled the reliable transport of bits; cloud-native systems enabled the orchestration of distributed computation. Today, the emergence of autonomous, learning-based systems introduces a new architectural challenge: intelligence is increasingly embedded directly into network control, computation, and decision-making, yet the Internet lacks a structural foundation for representing and exchanging meaning.
In this paper, we argue that cognition alone: pattern recognition, prediction, and optimization, is insufficient for the next generation of networked systems. As autonomous agents act across safety-critical and socio-technical domains, systems must not only compute and communicate, but also comprehend intent, context, and consequence. We introduce the concept of a Semantic Layer: a new architectural stratum that treats meaning as a first-class construct, enabling interpretive alignment, Semantic accountability, and intelligible autonomous behavior.
We show that this evolution leads naturally to a Syntactic-Semantic Internet. The syntactic stack continues to transport bits, packets, and workloads with speed and reliability, while a parallel Semantic stack transports meaning, grounding, and consequence. We describe the structure of this Semantic stack—Semantic communication, a Semantic substrate, and an emerging Agentic Web, and draw explicit architectural parallels to TCP/IP and the World Wide Web. Finally, we examine current industry efforts, identify critical architectural gaps, and outline the engineering challenges required to make Semantic interoperability a global, interoperable infrastructure.

 
\end{abstract}
	
\section{Introduction}
Modern networks are becoming cognitive. Through advances in machine learning, distributed control, and edge–cloud convergence, systems can now sense their environments, detect patterns, optimize behavior, and act autonomously. These capabilities mark a fundamental shift from static, rule-driven infrastructures to adaptive, learning-based systems.
Yet cognition, by itself, is insufficient for the systems we are now building \cite{shahraki2020survey}.

Cognitive systems operate primarily through statistical inference. They identify correlations, predict likely outcomes, and select actions that optimize learned objectives \cite{r1}. What they lack is the ability to determine why an action matters, whether it aligns with intent, or what consequences it may impose beyond local optimization. Two messages may be identical in structure and statistically similar in content, yet differ profoundly in meaning: one may represent a routine update, the other a safety-critical constraint. Without explicit Semantic grounding, a cognitive system has no architectural basis for distinguishing between them.

As autonomous systems scale and begin interacting with other autonomous systems, these limitations compound. Individually well-behaved agents may diverge over time as models adapt, assumptions shift, and local contexts evolve. Semantic drift accumulates \cite{r2}. Interpretations that were once aligned gradually fragment, leading to miscoordination, contradictory actions, and emergent failure modes. These failures are not computational bugs; they are architectural consequences of systems that lack shared structures for meaning.

A historical parallel clarifies the challenge. Before the advent of TCP/IP, early networks were sophisticated yet isolated. Each worked well internally, but interconnection failed because there was no shared architecture for addressing, routing, and transport. The problem was not intelligence or capability, but the absence of universally agreed-upon constructs. Cognitive systems today occupy a similar position. Without a structural layer for representing, aligning, and exchanging meaning, autonomous systems remain islands of intelligence rather than participants in coherent multi-agent ecosystems
Cognition enables prediction but not interpretation, behavior but not understanding, action but not explanation \cite{dorri2018multi}. Emerging autonomous infrastructures require more. They require comprehension: the ability to situate information within shared intent, contextual constraints, and anticipated consequences. Achieving this shift demands an architectural response, not merely better models \cite{r3}. That response is the Semantic Layer.

The Semantic layer treats meaning as a first-class engineering construct. By explicitly modeling intent, context, constraints, and anticipated consequences, it enables networks not only to transport data, but also to support shared understanding, interpretive alignment, and accountability. In doing so, the Semantic layer provides a structural foundation for autonomous collaboration across heterogeneous systems, application domains, and temporal scales.

Within the Semantic layer, Semantic communication constitutes the most fundamental capability. Unlike conventional communication paradigms that prioritize syntactic accuracy and bit-level fidelity, Semantic communication focuses on interpretive fidelity, i.e., whether the task-relevant Semantic content of information is preserved at the receiver \cite{r4}. Through task- and intent-aware representations, explicit contextual grounding, and structured handling of uncertainty, Semantic communication allows systems to maintain consistent understanding even under bandwidth constraints, model heterogeneity, and dynamically evolving environments. The introduction of Semantic communication therefore enables the Semantic layer to move from a conceptual abstraction toward an implementable architectural component, forming the basis for higher-level functions such as Semantic alignment, Semantic routing, and multi-agent coordination \cite{r6}.

\section{Semantic Layer in Syntactic–Semantic Internet}
\begin{figure*}[ht] 
    \centering
    \includegraphics[width=\textwidth]{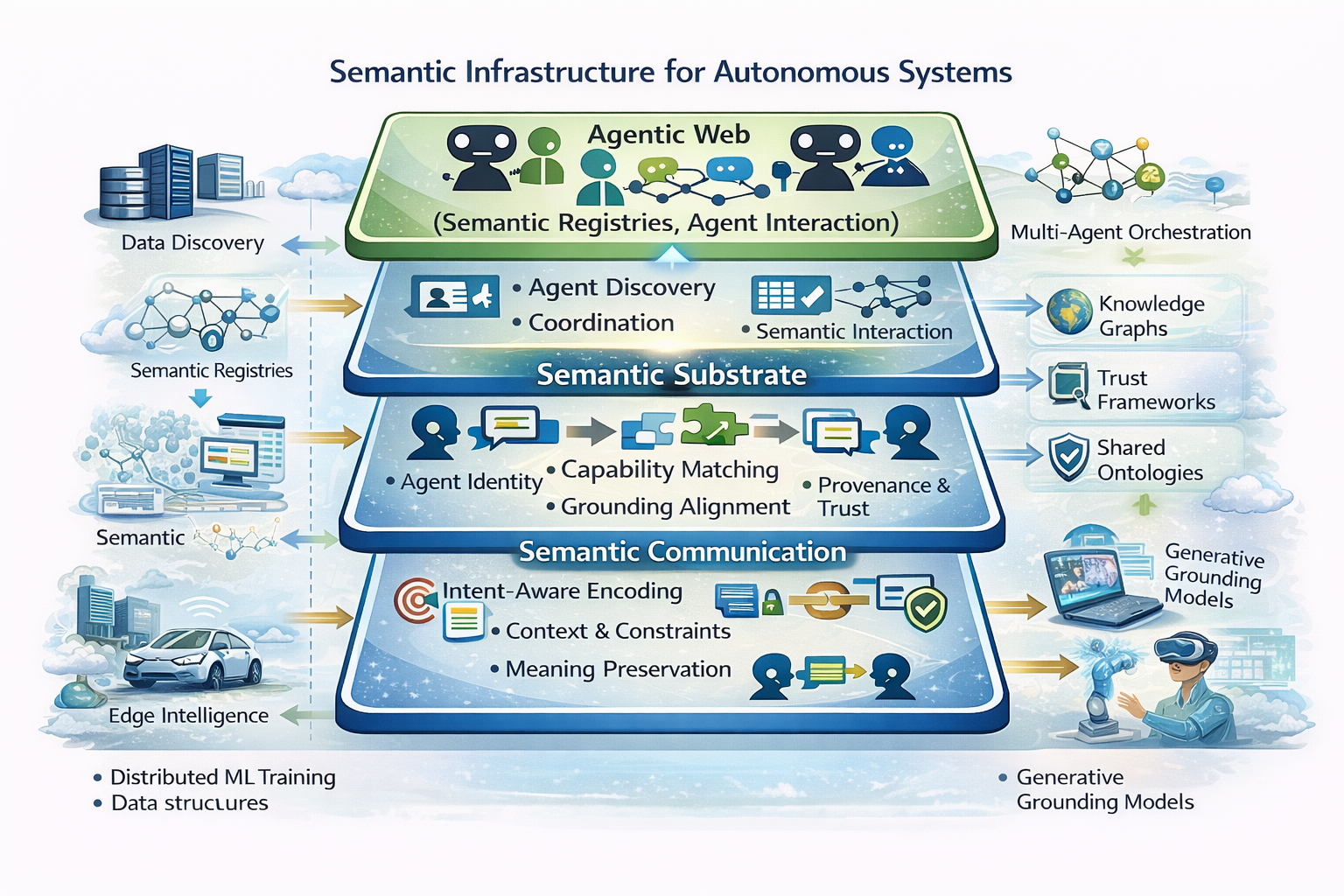}  
    \caption{Semantic infrastructure for autonomous systems.} 
    \label{fig:1} 
\end{figure*}

\subsection{From Data to Meaningful Action}
To support comprehension at scale, networks must carry not only information but the significance of that information. Just as packets gave structure to the movement of bits, and workloads gave structure to the movement of computation, the Semantic Layer gives structure to the movement of meaning. 

At its foundation, the Semantic Layer grounds information through contextual frames that capture the circumstances in which a message was created, the intent behind it, the constraints under which it should be interpreted, and the anticipated impact of acting upon it. These frames allow systems to distinguish superficially similar events that diverge fundamentally in meaning. They bind signals to concepts, concepts to intent, and intent to consequence.

Semantic communication evolves from preserving syntactic fidelity to preserving Semantic fidelity. Instead of transmitting every bit, systems transmit what is meaningful for a task; instead of minimizing uncertainty as noise, they represent uncertainty as an essential attribute of interpretation; instead of attempting to reconstruct data precisely, they attempt to preserve understanding. Information becomes task-relative, intent-relative, and consequence-aware \cite{shi2021new}.

The Semantic Layer also enables systems to negotiate meaning. When interpretive assumptions diverge or when information lacks sufficient grounding, systems must request clarification, renegotiate context, or decline to act. These behaviors are enabled not by cognitive heuristics but by structures that support interpretive alignment. The Semantic Layer therefore enforces the conditions under which comprehension can occur.

Finally, the Semantic Layer embeds the rationale behind actions. Cognitive models may output an action or a result without explanation. Semantic systems must expose why they acted: the intent, assumptions, grounding, provenance, and anticipated consequences behind each decision. This transparency transforms intelligent behavior into intelligible behavior \cite{zhong2025benchmark}.
\subsection{From Semantic Concept to Semantic Infrastructure}
Concepts alone do not scale; infrastructure does. The Internet’s success is rooted in its architecture: abstractions, layers, and protocols that operationalized universal connectivity. The Semantic Layer requires the same discipline. To transform meaning from a localized property of individual models into a globally exchangeable resource, we must engineer a Semantic infrastructure composed of three tiers: Semantic communication, the Semantic substrate, and the Agentic Web in Figure \ref{fig:1}.

\textbf{Semantic communication}: The Semantic communication layer redefines how information is represented and transmitted. Rather than treating data as anonymous symbols, it encodes each message within grounding frames that capture its intent, context, and expected consequences \cite{liu2025joint}. Representation becomes adaptive to the interpretive needs of the receiving agent. Messages emphasize critical Semantic features and elide irrelevant detail. This foundation mirrors the role of the physical and link layers in the syntactic stack, providing the channels through which meaningful information passes.
Errors are defined not as bit-level distortions but as divergences in meaning. Semantic communication provides the mechanisms by which meaning begins its movement across the system \cite{yang2022Semantic}.

\textbf{Semantic substrate}: Above Semantic communication lies the Semantic substrate: the architectural equivalent of TCP/IP for meaning. The substrate establishes the universal structures necessary for meaning to move across heterogeneous systems. It defines identities for agents, the contexts within which interactions occur, and the envelopes that carry meaning across interpretive boundaries. In this substrate, meaning is not simply passed from one actor to another; it is negotiated, aligned, stabilized, and attested.
Routing becomes an interpretive act \cite{r7}. Instead of forwarding messages based on topology, the Semantic substrate forwards meaning based on conceptual compatibility, capability, grounding alignment, and trust provenance. Systems become reachable not because they occupy a location in a network but because they possess the competence to interpret meaning correctly. Provenance is no longer metadata; it is a fundamental component of meaning. Drift detection becomes routine, enabling the substrate to identify when previously aligned interpretive structures begin to diverge \cite{lu2023Semantics}.

\textbf{Agentic Web}: Above the substrate lies the Agentic Web, which serves the same role for autonomous systems that the World Wide Web served for human users. It provides naming, discovery, interaction, and coordination across a landscape of autonomous agents. Just as the Web provided URLs, HTTP, HTML, and browsers that allowed humans to navigate information spaces, the Agentic Web provides Semantic registries, interaction protocols, shared grounding catalogs, and agentic runtimes that allow autonomous systems to navigate meaning spaces. It becomes the environment within which understanding is shared, negotiated, and acted upon.
These three tiers are intentionally minimal abstractions, designed to permit heterogeneity above and below \cite{abuelsaad2024agent}.

\section{Syntactic-Semantic Internet Architecture}

\begin{figure*}[ht] 
    \centering
    \includegraphics[width=\textwidth]{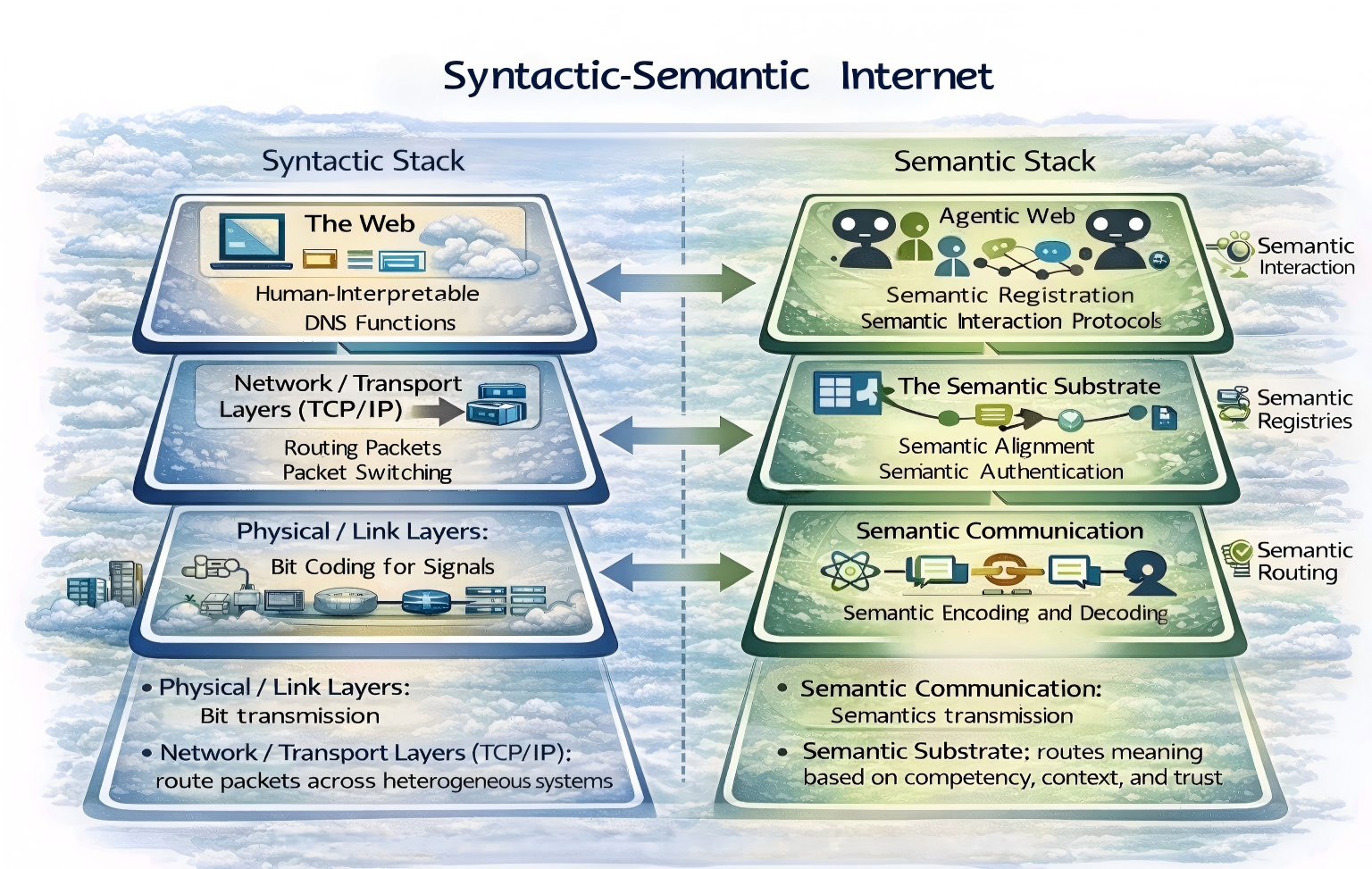}  
    \caption{Syntactic - Semantic Internet.} 
    \label{fig:2} 
\end{figure*}

Having defined the structure of the Semantic Layer, we now examine how it integrates with the Internet we have today. This integration must follow the core architectural principles that made the original Internet succeed: universality, layering, extensibility, and decoupling of internal implementation from external interoperability. The goal is not to replace the existing Internet but to extend it with a parallel Semantic architecture, precisely as transformative as the introduction of TCP/IP decades ago.

The Internet’s success stemmed from a simple insight: interconnection depends not on uniform internal design but on shared external structures. TCP/IP provided exactly this minimal universal envelope: addressing, routing, and transport Semantics, allowing radically different networks to communicate. The Web extended this with a universal naming, retrieval, and rendering model for information. Neither forced uniformity inside systems. Both provided the compatibility required for systems to interoperate.

Today, Semantic interoperability faces the same challenge that packet interoperability once did. Cognitive systems can generate insights, but these insights remain trapped inside local models, proprietary formats, or domain-specific ontologies. Without a universal Semantic envelope, analogous to packets, there is no architecture capable of supporting global, multi-agent, multi-domain comprehension.

The Semantic Layer therefore introduces not a new network but a parallel network layered atop the syntactic Internet. The two stacks operate together:
\begin{itemize}
\item the syntactic stack moves bits, packets, and workloads
\item the Semantic stack moves meaning, grounding, and consequence
\end{itemize}
This Syntactic-Semantic model is foundational and inevitable.

\subsection{Architectural Parallels Between the Two Stacks}
As shown in Figure \ref{fig:2}, The syntactic stack is built around three central abstractions:

1.Physical/Link Layers: encode bits into signals

2.Network/Transport Layers (TCP/IP): route packets across heterogeneous systems

3.The Web: provides human-interpretable information spaces

These abstractions energized the Internet’s scalability by separating concerns: bit transport, packet exchange, and information access.

The Semantic stack introduces corresponding abstractions:

1.Semantic Communication: encodes meaning relative to grounding frames

2.The Semantic Substrate: routes meaning based on competency, context, and trust

3.The Agentic Web: provides shared environments for agentic cooperation and Semantic interaction

These Semantic abstractions preserve the same layering discipline, but operate in a different domain: meaning rather than bits.
The stacks are therefore architectural counterparts, not alternatives.

\subsection{Why a Syntactic-Semantic Model Is Necessary}
The syntactic Internet excels at transporting bits and workloads with extraordinary speed and reliability. But bits do not carry meaning. Packets are indifferent to intent. TCP cannot verify interpretive alignment. HTTP cannot guarantee that two systems share grounding, rationale, or consequence models. The syntactic stack is blind to Semantics, not by failure but by design.

Autonomous systems, however, cannot act blindly. They must know:
\begin{itemize}
\item what a message means
\item under what assumptions it was produced
\item the obligations it carries
\item the consequences its interpretation may produce
\item the provenance and trustworthiness of its Semantic lineage
\end{itemize}
These requirements cannot be met by today’s syntactic architecture. They require a distinct Semantic architecture running alongside it. A Syntactic-Semantic future is therefore not optional. It is the architectural consequence of introducing comprehension into global systems.

\subsection{How the Two Stacks Operate Together}
The Syntactic-Semantic Internet functions through cooperative layering, not replacement.
\begin{itemize}
\item Syntactic channels provide connectivity, addressing, flow control, and packet integrity.
\item Semantic channels provide grounding synchronization, interpretive alignment, consequence management, and provenance guarantees.
\end{itemize}
A Semantic envelope is carried inside a packet but interpreted through the Semantic substrate. The packet ensures delivery; the Semantic envelope ensures understanding. Syntactic errors are corrected through retransmission; Semantic errors trigger renegotiation of meaning, realignment of grounding, or requests for clarification.

Semantic flow control differs from syntactic flow control: instead of regulating packet volumes, it regulates interpretive load, alignment stability, and the Semantic coherence of multi-agent exchanges.

DNS provides a naming system for resources; Semantic registries provide grounding authorities for meaning. These registries are not DNS, they do not map names to IP addresses, but they play an analogous role, mapping agents to competencies, grounding frames, and Semantic capabilities.

Over time, Semantic substrates, registries, and runtimes will become the invisible machinery of comprehension, just as routers and caches became the invisible machinery of connectivity.

\subsection{Syntactic-Semantic Internet Architecture}
To make this architectural evolution explicit, we summarize the parallel roles in Table \ref{tab:syntactic-Semantic-internet}.

\begin{table*}[htbp]
\centering
\caption{The Syntactic-Semantic Internet Architecture}
\label{tab:syntactic-Semantic-internet}
\begin{tabular}{|l|p{5cm}|p{6cm}|}
\hline
\textbf{Dimension} & \textbf{Syntactic Internet} & \textbf{Semantic Internet} \\
\hline
\textbf{Foundational Purpose} & Transport bits & Transport meaning \\
\hline
\textbf{Core Abstraction} & Packet & Semantic Envelope \\
\hline
\textbf{Representation Focus} & Symbols and data structures & Intent, context, grounding, consequence \\
\hline
\textbf{Error Model} & Bit-level fidelity & Interpretive fidelity \\
\hline
\textbf{Routing Principle} & Topology-driven packet forwarding & Capability-, context-, and intent-driven meaning routing \\
\hline
\textbf{Identity} & IP addresses & Agent identities with grounding and policy constraints \\
\hline
\textbf{Interoperability Basis} & Shared syntax and standards & Shared interpretation, alignment, and Semantic agreement \\
\hline
\textbf{Trust Model} & Channel reliability, encryption, authentication & Provenance, rationale, Semantic attestation, accountability \\
\hline
\textbf{Application Environment} & Web (URLs, HTTP, HTML, browsers) & Agentic Web (Semantic registries, agent runtimes, A2A protocols) \\
\hline
\textbf{Ultimate Goal} & Connectivity and data exchange & Coherence, comprehension, and accountable autonomy \\
\hline
\end{tabular}
\end{table*}

\subsection{Meaning in Motion: Toward Semantic Resilience}
Just as bits face noise, loss, and corruption as they traverse networks, meanings face Semantic drift, misalignment, or loss of grounding as they traverse agents, domains, and time. Packets are largely stateless; Semantic envelopes are inherently stateful. They carry assumptions, intent, provenance, rationale, obligations, and policy bindings.

Preserving meaning therefore requires architectural mechanisms beyond reliable transport:
\begin{itemize}
\item Semantic error detection
\item interpretive integrity checks
\item grounding preservation
\item Semantic retransmission
\item alignment protocols
\item consequence-sensitivity mechanisms
\end{itemize}
If the syntactic Internet engineered resilience for bits, the Semantic Internet must engineer resilience for understanding.

\subsection{Industry Initiatives and the Architectural Gaps That Remain}
Across industry, academia, and standards communities, momentum toward Semantic interoperability and agentic systems is unmistakable. Progress is occurring simultaneously in Semantic communication, agent runtimes, capability negotiation, and multi-agent coordination. Yet despite this activity, the current landscape mirrors the Internet before TCP/IP or the Web: innovation is advancing within individual layers, but without the architectural coherence required to form a global Semantic infrastructure.

At the foundation, research in Semantic communication is redefining how information should be encoded when the objective is comprehension rather than bitwise reconstruction. Efforts within UK’s Communications Hubs, Europe’s Hexa-X program, the U.S. NSF’s RINGS and VINES initiatives, and parallel programs across Asia explore task-aware representations, meaning-preserving encodings, and shared generative grounding \cite{r5}. These efforts illuminate what a Semantic counterpart to the physical and link layers might resemble, but they largely remain vertically isolated, without integration into Semantic routing, alignment, or agentic negotiation.

Above this, multiple initiatives are probing aspects of a Semantic substrate. Frameworks such as Anthropic’s MCP, Google’s A2A, IBM’s ACP, Cisco’s Internet of Agents, and MIT’s NANDA examine how autonomous systems discover one another, advertise capabilities, negotiate assumptions, and coordinate tasks. Each contributes essential primitives. None, however, yet constitutes the Semantic equivalent of TCP/IP: a minimal, universal substrate that allows meaning to move interoperably across heterogeneous agent ecosystems with shared guarantees of grounding, provenance, and interpretive fidelity.

At the application layer, early elements of an Agentic Web are beginning to emerge. Agent registries, discovery services, orchestration frameworks, and trust scaffolding are taking shape across open-source communities and industry consortia. These systems provide important local coherence, but they remain ecosystem-bound. What is still missing is a global Semantic registry, the true analogue of DNS, capable of resolving agent identities across domains and mapping them to competencies, grounding constraints, trust anchors, and policy contexts with Internet-scale universality.

Seen together, the pattern is clear. The components of a Semantic stack are advancing, but they are advancing in isolation. The industry is progressing within the Semantic stack, but not yet as a Semantic stack. What is absent is the architectural connective tissue: universal Semantic envelopes, cross-domain Semantic routing, and globally resolvable grounding and provenance.

This is precisely where the Syntactic-Semantic Internet becomes essential. Just as packet networking required a universal syntactic envelope and the Web required universal naming, Semantic interoperability requires a unifying architecture that binds communication, alignment, provenance, and interpretation into a coherent whole. Without this shared foundation, today’s agentic systems will remain powerful but fragmented. With it, they can converge into the next global substrate: an Internet capable not only of transporting data, but of transporting understanding.

\section{Engineering Challenges and the Road Ahead}
Engineering the Semantic Layer requires solving challenges far more intricate than those faced during the evolution of the syntactic Internet. TCP/IP dealt with bit-level corruption, congestion collapse, heterogeneous link technologies, and global routing. The Semantic Internet inherits all these challenges and adds a new class of problems, those concerned not with \textit{information} but with \textit{interpretation}. Meaning, unlike bits, is inherently contextual, dynamic, and stateful. Preserving it across distributed agents therefore demands new forms of architectural discipline, new protocol categories, and new operational mechanisms.

A central challenge is \textit{Semantic routing}: determining how \textit{meaning} should flow through a population of agents that differ in competence, grounding models, obligations, and trust relationships. Packet routing forwards symbols based on topology and cost; Semantic routing must forward conceptual payloads to agents capable of interpreting their assumptions, constraints, and embedded intent. A Semantic envelope cannot be delivered indiscriminately; it must be matched to an agent with sufficient grounding and interpretive capability. This matching requires metadata describing conceptual domains, trust anchors, contextual lineage, and interpretive relevance. Designing routing protocols that honor these constraints represents one of the most profound architectural challenges of the Semantic Internet.

Equally foundational is the development of \textit{alignment and grounding protocols}, the mechanisms that allow agents to negotiate or re-establish shared understanding. Over time, even well-aligned agents diverge, models drift, ontologies fragment, assumptions evolve, and interpretive frames adapt to local contexts. The Semantic Layer must make such drift detectable, measurable, and repairable. Alignment protocols must enable agents to query one another’s grounding, negotiate Semantic differences, resolve conflicts, and refuse actions when meaning cannot be safely reconstructed. In this sense, alignment behaves like Semantic flow control: not regulating packet rates but regulating the coherence and stability of interpretive exchange. Semantic alignment becomes the stabilizing mechanism that preserves coherence in multi-agent ecosystems, much as transport protocols stabilized packet flow across the early Internet.

\textit{Trust and provenance} must also be elevated from syntactic protection to Semantic assurance. In today’s networks, trust is rooted in encryption, authentication, and channel integrity. In Semantic systems, trust must additionally account for the origin, rationale, grounding, and consequence of meaning. Each Semantic envelope must carry a verifiable lineage: who produced it, under what assumptions, with which model versions, subject to which obligations, and transformed by which interpretive steps. Without such provenance, autonomous systems cannot determine whether a meaning is valid, safe, compliant, or aligned with collective norms.

\textit{Operational management}, the OAM of Semantic systems, requires new metrics and observability mechanisms. Traditional networks measure latency, throughput, jitter, and packet loss. Semantic networks must measure interpretive fidelity: the degree to which received meaning reflects sender intent. This requires new observability mechanisms capable of tracking Semantic drift, measuring grounding discrepancies, quantifying ambiguity, and evaluating interpretive uncertainty. Semantic retransmission may be needed not because packets were lost, but because understanding was.

Above all, \textit{standards} must evolve to support these new constructs. Just as the syntactic Internet required standardized packet formats, addressing schemes, transport Semantics, and application-layer protocols, the Semantic Internet requires Semantic envelope specifications, grounding registries, agent-to-agent protocols, provenance formats, capability directories, and Semantic routing rules. These standards will emerge only through open governance, cross-disciplinary consensus, and iterative experimentation, echoing the design culture that shaped TCP/IP.

\textit{Interoperability} testing will also take on a fundamentally new role. Instead of simply verifying whether two systems can exchange packets, future interoperability events must determine whether they can \textit{exchange meaning}: whether agents interpret concepts consistently, negotiate grounding effectively, detect and repair drift, propagate provenance securely, and act according to shared norms. These tests will define the practical boundaries of Semantic interoperability, just as early TCP/IP bake-offs defined the boundaries of syntactic interoperability.

Together, these challenges signal not merely the extension of existing architectures but the emergence of a fundamentally new one: a layer that binds meaning to action, aligns autonomous systems across domains, and enables comprehension to move across networks with the same reliability that packets move today. The engineering of this layer will shape not only how intelligent systems communicate, but also how they reason, cooperate, and remain accountable in a world where understanding itself moves across networks.

\section{Conclusion}
Across its history, the Internet has advanced by introducing architectural layers that abstract complexity while preserving interoperability. Packets stabilized global communication. Cloud-native abstractions stabilized distributed computation. Today, autonomous agents are stabilizing distributed intelligence, but intelligence alone is not sufficient for systems that increasingly act on the world.

As autonomy expands into safety-critical, economic, and socio-technical domains, systems must do more than compute and communicate. They must comprehend. Comprehension requires the ability to represent intent, situate information within context, reason about consequence, and remain accountable as meaning propagates across agents, domains, and time.

This paper has argued that these requirements cannot be met by extending the existing Internet stack alone. Instead, they necessitate a new architectural stratum: the Semantic Layer, which treats meaning itself as a first-class engineering construct. From this follows a natural and inevitable outcome: the Syntactic-Semantic Internet. The syntactic stack continues to ensure performance, reachability, and packet-level resilience. In parallel, the Semantic stack stabilizes the movement of meaning through grounding, provenance, interpretive alignment, and consequence awareness.

The two stacks are not competing paradigms but complementary infrastructures. One moves bits with fidelity; the other moves understanding with accountability. Together, they form the foundation for autonomous systems that can be trusted, reasoned about, and governed at planetary scale.

But meaning in motion introduces new vulnerabilities. Just as packets face loss and corruption, meaning faces Semantic drift, misalignment, and loss of grounding as it traverses agents and evolves over time. Semantic correctness cannot be defined by equivalence alone; it must be preserved as a stateful object carrying intent, assumptions, provenance, and obligation. Ensuring this demands new architectural mechanisms for Semantic error detection, alignment repair, grounding recovery, and consequence-aware renegotiation.

If the syntactic Internet engineered resilience for bits, the Semantic Internet must engineer resilience for understanding.

The Syntactic-Semantic Internet is therefore not a speculative vision but an architectural necessity. It represents the next phase in the Internet’s evolution: from connectivity, to computation, to comprehension. The design and standardization of this Semantic infrastructure will shape how autonomous systems cooperate, how responsibility is assigned, and how meaning itself can be safely exchanged in an increasingly agentic world.

 \bibliographystyle{IEEEtran}
	\bibliography{ref}
    \vspace{-10pt}
\end{document}